\documentstyle[aps,prc,preprint,tighten,epsf,dina4]{revtex}
\newcommand{\matrx}[4]{ \left( \begin{array}{*{2}{c}}
                                      #1 & #2 \\
                                      #3 & #4 
                                    \end{array} \right) }
\newcommand{\zweivek}[2]{ \left( \begin{array}{c} #1 \\ #2 
                               \end{array} \right) }
\newcommand{\braket}[2]{\left\langle{#1}\right.\left|{#2}\right\rangle}
\newcommand{\bramket}[3]{\left\langle\,{#1}\,\left|\,{#2}\,
            \right|\,{#3}\,\right\rangle}
\newcommand{\intp}[1]{\int\frac{d^{3}{#1}}{(2\pi)^{3}}}
\newcommand{\intpp}[1]{\int\frac{d^{4}{#1}}{(2\pi)^{4}}}
\newcommand{\bqn}{\begin{eqnarray}}
\newcommand{\eqn}{\end{eqnarray}}
\newcommand{\nn}{\nonumber\\}
\newcommand{\ov}{\overline}
\newcommand{\eps}{\varepsilon}
\newcommand{\half}{\frac{1}{2}}

\newcommand{\dagg}{^\dagger}
\begin{document}
\title{Electromagnetic N-$\Delta$ transition form factors in a
  covariant quark-diquark model}
\author{V. Keiner}
\address{Institut f\"ur Theoretische Kernphysik,\\
         Universit\"at Bonn, Nussallee 14-16, D-53115 Bonn, FRG}
\date{October 10, 1996}
\preprint{\vbox{Bonn TK-96-28}}

\maketitle

\begin{abstract}
The electromagnetic N-$\Delta$ transition form factors are calculated
in the framework of a formally covariant constituent diquark model. 
As a spin-$\frac{3}{2}$ particle the $\Delta$ is assumed to be a bound state
of a quark and an axial-vector diquark. The wave function is obtained from 
a diquark-quark Salpeter equation with an instantaneous quark exchange 
potential. The three transition form factors are calculated for momentum 
transfers squared from the pseudothreshold $(M_\Delta-M_N)^2$ up to 
$-2 \; (\mbox{GeV/c})^2$. The magnetic form factor is in 
qualitative agreement with experiment. We find very interesting results for
the ratios $E2/M1$ and $C2/M1$. 
\end{abstract}

\vspace{1cm}
PACS numbers: 14.20.Dh, 11.10.St, 12.39.Ki, 13.40.Gp, 14.20.Gk \\

\vspace{4cm} 
\noindent
e-mail: {\em keiner@pythia.itkp.uni-bonn.de} \\
Tel.: +49 (0)228 73 2377 \\
Fax: +49 (0)228 73 3728 \\ 

\newpage

\section{Introduction}
The study of the electromagnetic N-$\Delta$ transition form factors is
of extreme current interest \cite{baryon}. While the transition is dominated
by the magnetic $M_{1+}$ amplitude in the resonance region the contribution
from the electric $E_{1+}$ and $S_{1+}$ amplitudes is suppressed and even
zero in spherically symmetric quark models. Their ratio to the $M_{1+}$, 
however, reveals many details about the structure of the (excited) nucleon.
In a classical picture, a non-zero value of the quadrupole ratios
$E2/M1$ and $C2/M1$ indicates an oblate deformation of the $\Delta$.
$SU(6)$ symmetric quark models can account for this behaviour
by introducing tensor forces between the quarks, thus leading to a
configuration mixing of s and d states. This then also results 
in a non-vanishing electric form factor of the neutron. See 
Refs. \cite{drechsel,bernstein} for recent overviews concerning the
ratio $E2/M1$. When thinking of a non-symmetric 
$\Delta$-resonance the idea of introducing diquarks as correlated 
two-quark subsystems seems most striking. Since a quark-diquark model is able
to explain in a natural way the negative mean square charge radius of the 
neutron a clarification of the experimental situation of the N-$\Delta$
transition seems to be at hand. \\
The aim of this paper is not only to test a relativistic quark-diquark model
introduced in earlier works \cite{kea,keb}. Clearly, the nucleon is not only 
a system of a quark and a point-like scalar or axial-vector diquark
(called v-diquark in the following), see \cite{anselmino,leinweber} for
a discussion. Even less is the $\Delta$ a bound state of only a 
v-diquark and a quark. Nevertheless, it is worthwhile
to explore the results obtained by a
pure quark-diquark picture. A similar approach \cite{kroll} using light-cone
wave functions could account for a variety of experimental data at higher
energies. The results may qualitatively ask for strong quark-quark 
correlations in three-quark models. Thereby, the formally covariant
character of our model facilitates the discussion and justifies the 
calculation of the form factors up to intermediate momentum transfers.  
Apart from this we use the opportunity to list some interesting formulae
not found in the literature concerning the N-$\Delta$ transition 
form factors and transition currents. \\
The fundamental relativistic equation describing a two-body bound
state is the Bethe-Salpeter equation. 
Adopting the idea of a quark exchange interaction
from previous works \cite{reinhardt,buck,huang,meyer} we deduced a
pair of coupled  
Salpeter equations in the instantaneous approximation \cite{kea,keb}.
In this paper we apply this formalism also to the $\Delta$-resonance
with spin $\frac{3}{2}$.
Here, only the v-diquark component contributes. Thus, the calculation of
the N-$\Delta$ transition form factors projects out the v-diquark component 
of the nucleon, as far as scalar--v-diquark transitions are neglected
\cite{kroll,anselkroll}. We will see that the inclusion of these gives 
a surprisingly better agreement with the experimental data, especially 
of the magnetic neutron form factor. \\
This paper is structured as follows. In Sec. \ref{secmodel} we extend 
our quark-diquark model to the $\Delta$, thus obtaining the 
$\Delta$ Salpeter amplitude. In Sec. \ref{seccurrent} the calculation
of the transition currents is outlined. An interesting threshold relation is 
derived. Sec. \ref{secff} then shows how the transition form factors are 
obtained from the currents. In Sec. \ref{secresults} we present the results
and compare with the experimental data. Finally, in Sec. \ref{secsumm}
a summary is given.

\section{The model}
\label{secmodel}
We describe the nucleon as a relativistically bound state of a scalar
or v-diquark and a quark. The fundamental equation of this two-body
problem is the Bethe-Salpeter equation \cite{bethe}. Assuming an
instantaneous quark exchange interaction we derived a system of coupled
Salpeter equations. The details of the model are found in \cite{kea,keb}. 
In the rest frame of the nucleon we defined the Salpeter amplitude:
\bqn
\Psi_{(\mu)}(\vec p\,) & := & \gamma^0 \, \left. \int \frac{dp^0}
{2 \pi} \,  e^{i PX} \, \int d^4x \, e^{i px} \,
\bramket{0}{T \, \phi_{(\mu)}(x_1) \, \psi(x_2)}{P}  
\right|_{P=(M,\vec 0\,)} \; .
\eqn
The optional Lorentz index $\mu$ is to be applied only in the 
v-diquark channel. 
This index and all other indices are suppressed in the following. 
The amplitude $\Psi$ fulfills the quark-diquark Salpeter equation:
\bqn
({\cal H} \Psi)(\vec p\,) & = & M \, \Psi(\vec p\,) \nn
& = & \frac{\omega_1+\omega_2}{\omega_2} \, H_2(\vec p\,) \, \Psi(\vec p\,)
      +\frac{1}{2 \omega_1} \, \intp{p'} 
       W(\vec p, \vec p\,') \, \Psi(\vec p\,') \; .
\eqn
Following the ideas of similar quark-diquark models 
\cite{reinhardt,buck,huang,meyer,ishii} the interaction
kernel is simply a quark exchange propagator (in the static approach):
\bqn
W(\vec p, \vec p\,') & \sim &  
g^2 \, \frac{1}{\omega_q^2} \, (-\vec\gamma(\vec p+\vec p\,')+m_q) \; ,
\eqn
with $\omega_q$ the energy of the exchanged quark and $g$ the quark-diquark
coupling parameter. Whereas in this picture 
the nucleon is a coupled system with a scalar and a v-diquark channel the
$\Delta$ is a bound state of only a v-diquark and a quark. Thus, the
quark-diquark Salpeter equation in Ref. \cite{keb} simplifies to:
\bqn \label{deltasalpeter}
M_\Delta \, \Psi_{\Delta \; S \; \alpha}^{[1]}(\vec p\,) & = & 
\frac{\omega_1+\omega_2}{\omega_2} \, H_2(\vec p\,) \,
\Psi_{\Delta \; S \; \alpha}^{[1]}(\vec p\,) \\
& & +\frac{1}{2 \omega_1} \, \intp{p'} \, (+{g_v^\Delta}^2) \, 
\frac{1}{\omega_q^2} \, 
\gamma_\beta^{[1]} \, (\vec \gamma (\vec p+\vec p\,')+m_q) \, 
(-{\gamma_\alpha^{[1]}}\dagg) \, \gamma^0 \,
\psi_{\Delta \; s \; \beta}^{[1]}(\vec p\,') \; . \nonumber
\eqn
The tensor rank $'[1]'$ of the amplitude indicates its vector character,
with $\alpha, \beta = 1 \dots 3$ in the $\Delta$ rest frame. $S$ is the 
z-component of the total spin. An additional Gaussian diquark form
factor \cite{kea} is suppressed.
Eq. (\ref{deltasalpeter}) is solved by expanding
$\Psi$ in a finite basis and using the Ritz variational principle.

\section{Current matrix elements}
\label{seccurrent}
\begin{figure}
\begin{center}
\leavevmode
\epsfxsize=0.60\textwidth
\epsffile{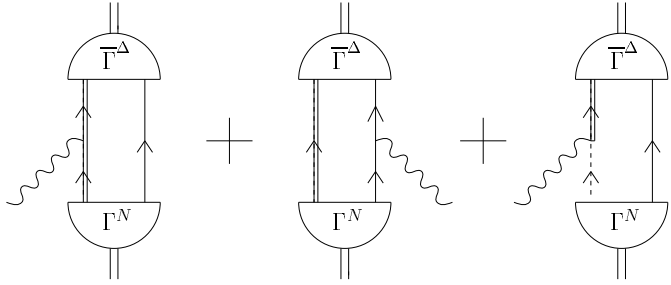}
\end{center}
\caption{The N-$\Delta$ transition current is the sum of the v-diquark
  current and the quark current. The third diagram is the
  scalar--v-diquark transition (see Eq. (\protect\ref{svtrans})). } 
\label{figcurrent}
\end{figure}

As in \cite{kea,keb} the electromagnetic N-$\Delta$ transition currents 
are calculated in the Mandelstam formalism \cite{mandelstam}. 
In a first step we only consider the first two diagrams of Fig. 
\ref{figcurrent}. 
Since the $\Delta$ is a pure v-diquark--quark state the N-$\Delta$ 
transition picks up only the nucleon's v-diquark channel. 
E.g. the quark current is: 
\bqn \label{qcurrent}
\lefteqn{
\bramket{\Delta(P' s')}{j_\mu^{quark}}{N(P s)} } \nn
& = & 
e_2 \intpp{p} \ov\Gamma_{P'}^{\Delta \; s'}(p') \,  
S_2^F(p'_2) \, \gamma_\mu \, S_2^F(p_2) \,
\Gamma_P^{N \; s}(p) \, \Delta_1^F(p_1)  \; .
\eqn
We recall the definition of the vertex as the amputated Salpeter amplitude:
\bqn
\Gamma(\vec p\,) & = & - i \, \intp{p'} \, W(\vec p, \vec p\,') \, 
\Psi(\vec p\,') \\
\ov \Gamma(\vec p\,) & = & - \Gamma\dagg(\vec p\,) \gamma^0 \; .
\eqn
As a spin-$\frac{3}{2}$ particle the $\Delta$ vertex transforms as
\bqn \label{rsboost}
\ov \Gamma_{P'}^{\Delta \; \mu}(p') & = & 
\ov \Gamma_{(M_\Delta,\vec 0\,)}^{\Delta \; \nu}(\Lambda^{-1}(p')) \,
\Lambda^\mu_{\;\;\nu} \, S_\Lambda^{-1} \; ,
\eqn
where $P'=\Lambda \; (M_\Delta,\vec 0\,)$. Including the flavour 
dependence the $\Delta$ amplitude reads
\bqn
\Psi_{\Delta \; s}(\vec p\,) & = &
{\cal N} \, \Psi_{\Delta \; s}^{[1]}(\vec p\,) \, 
\frac{1}{\sqrt{3}} \, 
\left( \sqrt{2} \, \chi_0^{[1]} + \chi_{+1}^{[1]} \right) \; ,
\eqn
with ${\cal N}$ such that the normalization according to the scalar 
product of Eq. (14) in \cite{kea} fulfills
\bqn
\braket{\Psi_\Delta}{\Psi_\Delta} & = & 2 \, M_\Delta \; .
\eqn
We then obtain for the N-$\Delta$ transition current:
\bqn \label{ndcurrent}
J_\mu(q^2) & := & \bramket{\Psi_\Delta}{j_\mu}{\Psi_N} =  
\frac{\sqrt{2}}{3} \, \left(
\bramket{\Psi_\Delta^{[1]}}{j_\mu^{diquark}}{\Psi_N^{[1]}}
- \bramket{\Psi_\Delta^{[1]}}{j_\mu^{quark}}{\Psi_N^{[1]}}
\right).
\eqn
As in the case of the elastic neutron current, the N-$\Delta$
transition current is sensitive to the difference of the quark and diquark
currents. Therefore we should not expect a better description of 
the N-$\Delta$ form factors than of the elastic neutron form factors
calculated in Ref. \cite{keb}. Finally, we want to state an interesting 
relation between the quark and diquark currents of Fig. \ref{figcurrent} 
at the pseudothreshold $q_s^2 := (M_\Delta-M_N)^2$.
We start by assuming the $\Delta$ vertex to transform like a spinor, 
thus dropping the Lorentz boost matrix $\Lambda^\mu_{\;\;\nu}$ 
in Eq. (\ref{rsboost}), and find similar to Eq. (33) of 
Ref. \cite{kea} (writing N$^*$ instead of $\Delta$):
\bqn \label{jqdnulla}
\lefteqn{
\left. \bramket{N^* \; P'}{j_0^{quark}}{N \; P}\right|_{q_s^2} 
=: j_0^{quark}(q_s^2) } \\
& = &
\intpp{p} \, 
\ov\Gamma_{(M_{N^*},\vec 0\,)}^{N^*}(\vec p\,') \,  
S_2^F(p'_2) \, \gamma^0 \, S_2^F(p_2) \, 
\Gamma_{(M_N,\vec 0\,)}^N(\vec p\,) \, 
\Delta_1^F(p_1)  \\
& = & \intp{p} \, (2 \omega_1) \, \ov \Psi^{N^*}(\vec p\,) \, 
\gamma^0 \, \Psi^N(\vec p\,) \\
& = & \intpp{p} \, 
\ov\Gamma_{(M_{N^*},\vec 0\,)}^{N^*}(\vec p\,') \,  
S_2^F(p_2) \, \Gamma_{(M_N,\vec 0\,)}^N(\vec p\,) \, 
\Delta_1^F(p_1) \, (p_1+p'_1)^0 \, \Delta_1^F(p'_1) \\
\label{jqdnullb}
& = & \left. \bramket{N^* \; P'}{j_0^{diquark}} 
{N \; P}\right|_{q_s^2} 
=: j_0^{diquark}(q_s^2) \; .
\eqn
To obtain the N-$\Delta$ transition current density
$J_0(q^2) := J_0^{\Delta N}(q^2) = 
\bramket{\Delta \; P'}{j_0}{N \; P}$, the
N-N$^*$ current $J_0^{N^* N}$ has to be 
multiplied with a kinematical factor arising from the Lorentz boost matrix
$\Lambda^\mu_{\; \nu}$ of the outgoing $\Delta$ amplitude (see
App. \ref{app}):
\bqn
J_0(q^2) & = & \sum_{i=1,2} \, f^i(q^2) \, J_0^{N^* N \; i}(q^2) \; , \\
\mbox{with } \;\; 
f^1(q^2) & = & \sqrt{\frac{2}{3}} \, \frac{q}{M_\Delta} \;,\;\;
f^2(q^2) = \frac{\sqrt{2}}{3} \, \Big( 1-\frac{P'^0}{M_\Delta}
\Big) \; ,
\eqn
where $i$ numbers the v-diquark channel components of the nucleon, 
see App. \ref{app}.
Especially, we have $f^i(q_s^2) = 0$, and 
$J_0^{N^* N}(q_s^2) = \frac{\sqrt{2}}{3} \, 
(j_0^{quark}(q_s^2)-j_0^{diquark}(q_s^2)) = 0$ 
(Eqs. (\ref{ndcurrent}) and (\ref{jqdnulla})-(\ref{jqdnullb})). 
So we obtain (with $q^3 = P' = q = |\vec q\,|$):  
\bqn 
\label{jnull}
J_0(q_s^2) & = & 0  \\
\label{dpjnull}
\left. \frac{d}{dP'}J_0(q^2) \right|_{q_s^2} & = & 0 \; .
\eqn 
Eq. (\ref{jnull}) of course also follows from current conservation 
$q^0 \, J_0 = - q^3 \, J_3 = 0$ at $q_s^2$.
We will see that these relations will guarantee a finite Coulomb form 
factor $G_C$ at $q_s^2$.
Similar to the electric neutron form factor vanishing at $q^2 = 0$, 
we will even find $G_C(q_s^2) = 0$.

\section{N-$\Delta$ transition form factors}
\label{secff}
In analogy to the usual Sachs decomposition of the elastic electromagnetic 
nucleon current the N-$\Delta$ transition current is 
expanded in terms of {\em three} independent covariant and 
gauge-invariant tensors $G_{\beta \mu}$ \cite{jones,devenish,weber}:
\bqn \label{jndtrans}
e \, J_\mu(q^2) := \bramket{\Delta(P' s')}{J_\mu}{N(P s)} & = & 
e \; \sqrt{\frac{2}{3}} \;\, \ov u_{s'}^\beta(P') \, J_{\beta \mu} 
\, u_s(P) \; , 
\eqn
with the decomposition
\bqn \label{jmudec}
J_{\beta \mu} & = & G_M(q^2) \, G_{\beta \mu}^M + G_E(q^2) \, G_{\beta
  \mu}^E + G_C(q^2) \, G_{\beta \mu}^C \; .
\eqn
$G_M, G_E$ and $G_C$ are the conventional magnetic dipole, electric
quadrupole and Coulomb quadrupole transition
form factors. The flavour factor in Eq. (\ref{jndtrans}) arises from
the normalization convention of Ref. \cite{jones}. The tensors in
Eq. (\ref{jmudec}) are:
\bqn
G_{\beta \mu}^M & = & -3 \, (M_\Delta+M_N) \, \eps_{\beta \mu}(\tilde P q) 
/(2 M_N Q^+) \\
G_{\beta \mu}^E & = & -G_{\beta \mu}^M - 6 \, (M_\Delta+M_N) \,
\gamma^5 \, \eps_{\beta \sigma}(\tilde P  q) \,
\eps_{\mu}^{\;\;\sigma}(P' q)
/(M_N \, \Delta(q^2)) \\ 
G_{\beta \mu}^C & = & -3 \, (M_\Delta+M_N) \, q_\beta \,
(q^2 \tilde P_\mu - (q \cdot \tilde P) \, q_\mu) \, \gamma^5/(M_N \,
\Delta(q^2)) \; ,  
\eqn
with
\bqn
\tilde P & = & \half \, (P+P') \\
\eps_{\beta \mu}(\tilde P q) & = &  
\eps_{\beta \mu \lambda \rho} \, \tilde P^\lambda q^\rho \\
Q^\pm & = & (M_\Delta \pm M_N)^2-q^2 \;,\;\;\;
\Delta(q^2) = Q^+ Q^-  \; . 
\eqn
Note that in the original Refs. \cite{jones,devenish} the $\gamma^5$ is
defined via $\tilde\gamma^5 = \gamma^0 \gamma^1 \gamma^2 \gamma^3$.
In the rest frame of the incoming nucleon, and choosing 
$q^3 = P' = q = |\vec q\,|$ we find 
(with $\gamma^5 = i \tilde\gamma^5$ and dropping the global $i$):
\bqn \label{gmec}
\ov u_{+\half}^\beta(P') \, G_{\beta 0}^M \, u_{+\half}(P) & = & 
\ov u_{+\half}^\beta(P') \, G_{\beta 0}^E \, u_{+\half}(P) = 0 \\
\ov u_{+\half}^\beta(P') \, G_{\beta 0}^C \, u_{+\half}(P) & = & 
-\sqrt{6} \, \frac{P'}{2 M_\Delta} \; g(q^2) \\
\ov u_{+\half}^\beta(P') \, G_{\beta +}^M \, u_{-\half}(P) & = &
\frac{\sqrt{3}}{2 \sqrt{2}} \; g(q^2) \\
\ov u_{+\half}^\beta(P') \, G_{\beta +}^E \, u_{-\half}(P) & = &
\frac{\sqrt{3}}{2 \sqrt{2}} \; g(q^2) \\ 
\ov u_{+\half}^\beta(P') \, G_{\beta +}^C \, u_{-\half}(P) & = &
\ov u_{+\frac{3}{2}}^\beta(P') \, G_{\beta +}^C \, u_{+\half}(P) = 0 \\
\ov u_{+\frac{3}{2}}^\beta(P') \, G_{\beta +}^M \, u_{+\half}(P) & = &
\frac{3}{2 \sqrt{2}} \; g(q^2) \\
\ov u_{+\frac{3}{2}}^\beta(P') \, G_{\beta +}^E \, u_{+\half}(P) & = &
-\frac{9}{2 \sqrt{2}} \; g(q^2) \; ,
\eqn
where we defined
\bqn
g(q^2) & := & \frac{M_\Delta + M_N}{\sqrt{2 M_N
    (P'^0+M_\Delta)}} \, P'  =  
\frac{M_\Delta+M_N}{2 M_N} \, \sqrt{Q^-} \; .  
\eqn
Inverting these equations then yields:
\bqn 
\label{gme}
\zweivek{G_M(q^2)}{G_E(q^2)} & = & 
\sqrt{\frac{3}{2}} \;
\frac{2 \, \sqrt{2}}{g(q^2)} \,
\matrx{\frac{\sqrt{3}}{4}}{\frac{1}{12}}
     {\frac{1}{4 \sqrt{3}}}{-\frac{1}{12}} 
\zweivek{J_+(q^2)}{J'_+(q^2)} \\
\label{gc}
G_C(q^2) & = & - \sqrt{\frac{3}{2}} \;
\frac{2 M_\Delta}{\sqrt{6} \, P' \, g(q^2)} \, 
J_0(q^2)   = 
- \sqrt{\frac{3}{2}} \;
\frac{4 M_\Delta M_N}{g(q^2) \, \sqrt{6 \, Q^+ \, Q^-}} \, 
J_0(q^2) \; , 
\eqn
where we wrote $J'_+$ for the spin flip current 
$\bramket{\Delta_{+\frac{3}{2}}}{J_+}{N_{+\half}}$, with 
$J_+ = \half (J_1 + i J_2)$. 
As expected, the charge density $J_0$ contributes to the Coulomb form factor
only, while $G_M$ and $G_E$ are related to the two spin flip currents 
via a mixing matrix. Note that $G_E$ is essentially the difference of
$J_+$ and $J'_+$.

\section{Results and discussion}
\label{secresults}
\begin{figure}
\begin{center} 
\leavevmode
\epsfxsize=0.65\textwidth
\epsffile{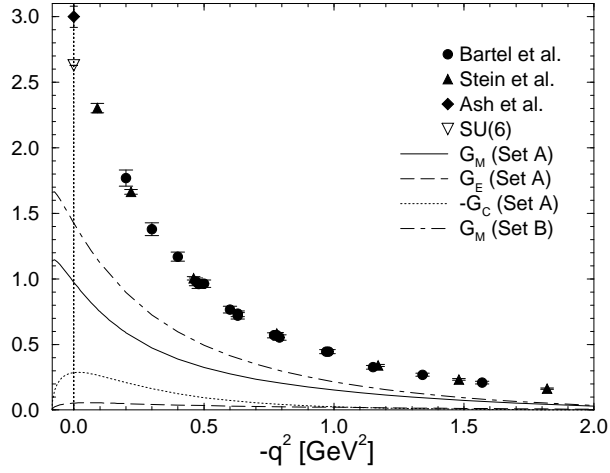}
\end{center}
\caption{The three N-$\Delta$ transition form factors. The data points
  show the measured magnetic form factor $G_M$
  \protect\cite{bartel,stein,ash}. The empty triangle follows
  from the equal-mass $SU(6)$ limit \protect\cite{jones}.
  The dash-dotted line is $G_M$ with scalar--v-diquark transitions
  included (Eq. (\protect\ref{svtrans})). } 
\label{ndeltaff}
\end{figure}

\begin{figure}
\begin{center} 
\leavevmode
\epsfxsize=0.65\textwidth
\epsffile{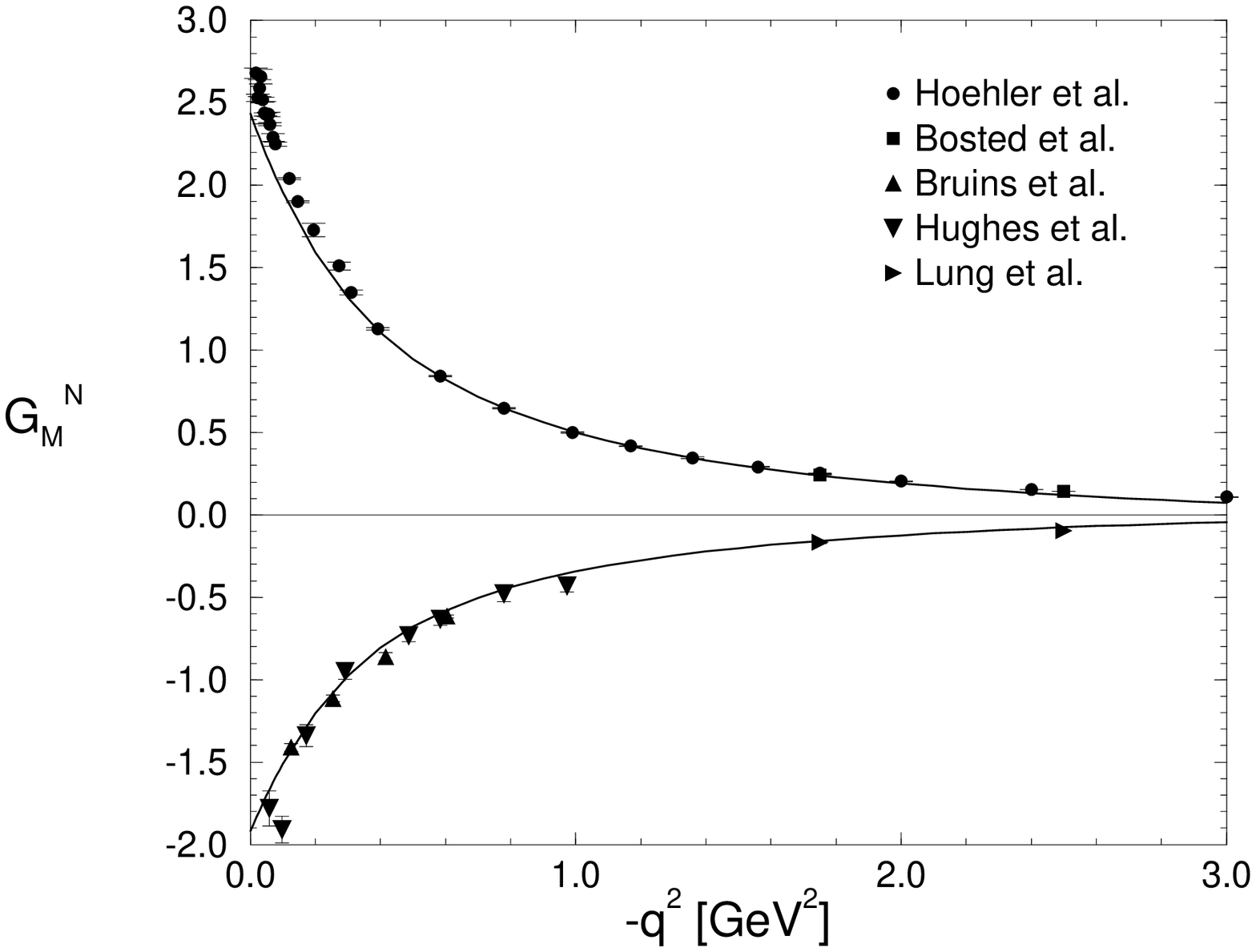}
\end{center}
\caption{The magnetic form factor of the proton (positive curve) 
  and of the neutron (negative curve) calculated with the parameter
  Set B of 
  Tab. \protect\ref{deltaparam}. For the experimental data see 
  the analysis of Ref. \protect\cite{ulf}. }
\label{nuclff}
\end{figure}

\begin{figure}
\begin{center} 
\leavevmode
\epsfxsize=0.65\textwidth
\epsffile{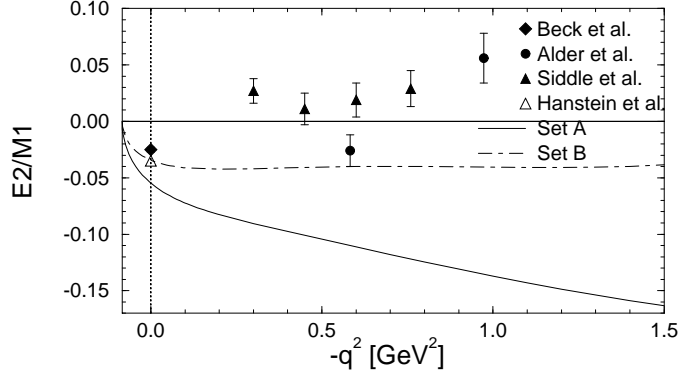}
\end{center}
\caption{The ratio $E2/M1$. The solid line results from the parameter
  Set A, the dash-dotted one from Set B. The experimental data are
  from Refs. \protect\cite{alder,siddle,beck}. The empty triangle is from
  the analysis of \protect\cite{drechsel}. } 
\label{e2ovm1}
\end{figure}

\begin{figure}
\begin{center} 
\leavevmode
\epsfxsize=0.65\textwidth
\epsffile{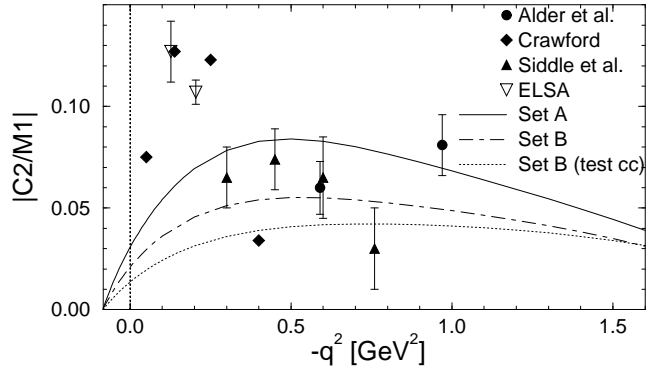}
\end{center}
\caption{The absolute value of the ratio $C2/M1$. The solid line
  results from the parameter 
  Set A, the dash-dotted one from Set B. The dotted line shows the
  ratio with $J_0$ calculated via 
  $J_0(q^2) = -\frac{q^3}{q^0} \, J_3(q^2)$. 
  The experimental data are
  from Refs. \protect\cite{alder,siddle,crawford,gothe}. }
\label{c2ovm1}
\end{figure}

\begin{figure}
\begin{center} 
\leavevmode
\epsfxsize=0.65\textwidth
\epsffile{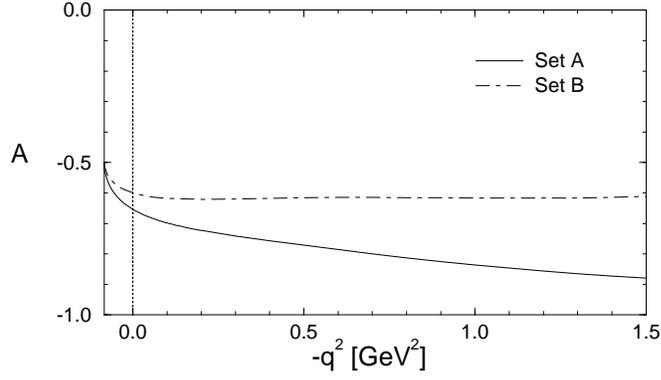}
\end{center}
\caption{The helicity asymmetry ratio $A$ (Eq. (\protect\ref{heliasy})) 
calculated with Set A (solid line) and with Set B (dash-dotted
line).}   
\label{asy}
\end{figure}

\begin{figure}
\begin{center} 
\leavevmode
\epsfxsize=0.65\textwidth
\epsffile{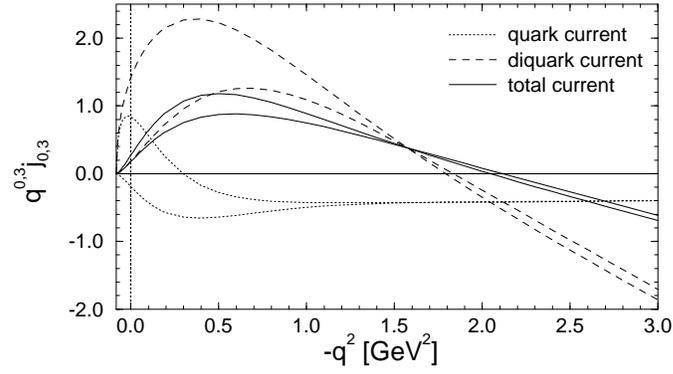}
\end{center}
\caption{The evolution of $q^0 j_0$ and $q^3 j_3$ of the first two currents
  in Fig. \protect\ref{figcurrent} and of the sum of both (solid
  line).}
\label{currcons}
\end{figure}

The parameters of the model are listed in Tab. \ref{deltaparam} (Set A). 
The parameters differ from those
of Ref. \cite{keb} in order to obtain a bound $\Delta$ without
introducing a confining potential. 
As the current study of the timelike nucleon electromagnetic form
factors shows, the bigger constituent (di-)quark masses also are
needed to obtain the correct threshold behaviour which is found to be
very sensitive to the masses.   
$\lambda$ is a parameter entering in the Gaussian diquark
form factor \cite{kea}. 
The nucleon parameters are fixed to obtain a best description
of the nucleon electromagnetic form factors. The 
resulting static properties are listed in Tab. 
\ref{tabresults} (Set A). The $q^2$ dependence of the four nucleon form
factors is nearly identical to that of Ref. \cite{keb}, therefore not
depicted in this article. However, we find a
little deterioration of $G_M^p(0)$ compared to 
Ref. \cite{keb}. It is interesting to find for the anomalous magnetic
moment of the v-diquark  
$\kappa = 1.1$, very near to the value of a point-like diquark 
($\kappa_0 = 1.0$).
In addition to Ref. \cite{keb} we have a new parameter $g_v^\Delta$ 
(see Eq. (\ref{deltasalpeter})) introduced in order to fix the $\Delta$ mass
at 1232 MeV. With $g_v^N=g_v^\Delta$ we would obtain $M_\Delta=1021$
MeV. This is a drawback of our model for the description of the 
static properties compared to other works as e.g. \cite{hanhart}. 
However, the correct 
description of the nucleon form factors up to $-3.0 \; \mbox{GeV}^2$
forced us to choose a scalar--v-diquark symmetric parameter set 
\cite{keb}, which fixes $g_v^N$.  Fig. \ref{ndeltaff} shows the three
calculated 
transition form factors for momentum transfers from the  
pseudothreshold $q_s^2$ up to $-2.0 \; \mbox{GeV}^2$. Note that
electron scattering 
experiments only access $q^2 \le 0$. The data points are the 
experimental $G_M$ \cite{bartel,stein,ash}. The empty triangle follows
from the equal-mass $SU(6)$ limit \cite{jones}.
Unfortunately, the calculated $G_M$
is a factor of 2.8 too low. This is not surprising in a model 
where the magnetic transition proceeds in the v-diquark channel alone.
This leads us to consider also transitions from a scalar to a
v-diquark (see the third diagram in Fig. \ref{figcurrent}),
which of course would contribute about equally in a three-quark
model. The coupling is analogous
to the $\pi-\omega$ transition \cite{anselkroll} and contains a
scalar--v-diquark coupling parameter $\kappa_{sv}$. We define it via
\bqn \label{svtrans}
j_\mu^{s \leftrightarrow v} & = & -2 \sqrt{2} \, i \, 
\frac{(1+\kappa_{sv})}{M_N} \, \eps_{\mu \nu \rho \lambda} \,  
\epsilon^\nu \, P'^\rho \, P^\lambda \; .
\eqn
In the rest frame of the 
nucleon and choosing $P'=P'_z$,
such a transition only affects the spin flip currents. So the
r.h.s. of Eq. (\ref{ndcurrent}) contains an additional term 
$+\frac{\sqrt{2}}{3 \sqrt{3}}
\bramket{\Psi_\Delta^{[1]}}{j_+^{diquark(v \leftrightarrow s)}}
{\Psi_N^{[0]}}$. 
Of course also the elastic nucleon form factors change. In order to find a 
best description of the magnetic form factors we choose Set B in 
Tab. \ref{deltaparam}. The resulting magnetic nucleon form factors are 
shown in Fig. \ref{nuclff}. The agreement with the experimental data
is indeed striking. With this Set B we obtain for the magnetic 
transition form factor the dash-dotted line in Fig. \ref{ndeltaff}. 
Here, we also find an improvement, with
the calculated curve still being too low, though.
Of special interest is the shape of $G_E$ and $G_C$. Both start
at zero at the pseudothreshold $q_s^2$ and have a maximum at about 
$q^2 = 0$. The threshold behaviour $G_C(q_s^2)=0$ results from 
Eqs. (\ref{jnull}) and (\ref{dpjnull}). 
As can be seen from Eq. (\ref{gme}) the electric form
factor itself is very sensitive to the difference of the two currents
$J_+(q^2)$ and $J'_+(q^2)$. It is straightforward to see that 
$J'_+(q^2) = \sqrt{3} \, J_+(q^2)$ for $q^2$ near to $q_s^2$. With 
$J_+(q_s^2) = J'_+(q_s^2) = 0$, this leads to $G_E(q_s^2)=0$. \\
In Fig. \ref{e2ovm1} we show the ratio of the multipoles 
$E2/M1 = E_{1+}/M_{1+} = - G_E(q^2)/G_M(q^2)$ \cite{jones,devenish}
compared to the experimental Re$(E_{1+} M_{1+}^*)/|M_{1+}|^2$ taken
from Refs. \cite{alder,siddle,beck}.  
The solid line corresponds to the parameter Set A. At $q^2=0$ 
the calculated ratio $E2/M1 = -5.5 \, \%$ is near to the experimental value
$E2/M1 = -2.5 \pm 0.2 \, \%$ \cite{beck}. Note, that the experimental
value still contains background effects. The recent analysis of
Ref. \cite{drechsel} gives $E2/M1 = -3.5 \, \%$ for the 'dressed'
$\Delta$ resonance alone. It is this number our calculation
has to be compared with.
For higher momentum transfers our curve decreases, similar to Ref. 
\cite{weber}, where a light-cone quark model is employed.  
This disagrees with the experimental data
from Refs. \cite{alder,siddle}. However, the experimental situation is 
not at all clear since large background effects hinder the extraction of 
$E2/M1$ from measured cross sections and make it strongly model
dependent \cite{drechsel,bernstein}. The dash-dotted curve is the same ratio
if we include scalar--v-diquark transitions according to 
Eq. (\ref{svtrans}), using Set B. The threshold value 
$E2/M1 = -3.4 \, \%$ is in astonishing agreement with the above value 
of Ref. \cite{drechsel}, and for higher  
$q^2$ the curve flattens. 
Fig. \ref{c2ovm1} shows the ratio \cite{kroll,devenish}
\bqn \label{coverm}
C2/M1 = S_{1+}/M_{1+} = - \frac{\sqrt{Q_+ Q_-}}{4 M_\Delta^2} 
\frac{G_C(q^2)}{G_M(q^2)} \; .
\eqn
The negative sign of $G_C$ leads to a positive ratio $C2/M1$ which
apparently contradicts the experimental data
\cite{alder,siddle,crawford,gothe}. Cardarelli et
al. \cite{cardarelli} also find $G_C < 0$, and for certain wave
functions also Kroll et al. \cite{kroll} find this behaviour. 
We strictly follow the definitions of Ref. \cite{devenish}. A positive
$J_0$ then leads straightforwardly via  
Eqs. (\ref{gc}), (\ref{coverm}) to $C2/M1 > 0$.
However, the absolute values of both Set A and Set B
are in satisfactory agreement with the experiment. We predict a
threshold value $C2/M1 = + 2.1 \, \%$ (Set B). The dotted line is the
ratio with the current density $J_0$ calculated via 
$J_0(q^2) = -\frac{q^3}{q^0} \, J_3(q^2)$ (also Set B). Thus, the
variance between the dash-dotted and the dotted curve reflects the
accuracy of our prediction due to the only partially conserved
current, see below. The dotted curve yields
$C2/M1 = + 1.4 \, \%$.  
Finally, Fig. \ref{asy} shows our prediction for the helicity
asymmetry ratio defined as 
\cite{kroll}
\bqn \label{heliasy}
A(q^2) & = & \frac{|A_\half|^2-|A_{\frac{3}{2}}|^2}
                  {|A_\half|^2+|A_{\frac{3}{2}}|^2} = 
-\half-3 \, \frac{G_M(q^2) G_E(q^2)-G_E^2(q^2)}
{G_M^2(q^2)+3 G_E^2(q^2)} \; ,
\eqn 
where $A_\half$ and $A_{\frac{3}{2}}$ are two of the three independent
electromagnetic helicity amplitudes \cite{kroll}. The ratio $A$
gives essentially the contribution of helicity nonconserving compared
to helicity conserving transitions. Symmetric three-quark models
should give a constant $A(q^2) \equiv -0.5$. For both Sets A and
B we see a similar deviation from this rule, with curve A decreasing
faster than curve B. Our results are very similar to those of
Ref. \cite{weber}. There, it is explicitely shown that a satisfactory
description of $G_M(q^2)$ and $E2/M1$ is mainly due to a correct
relativistic treatment. So we may conclude that apart from the assumed
quark-diquark structure our good results are also due to the
formally covariant Salpeter model. \\
At last we should mention an interesting result concerning current 
conservation. In the previous Refs. \cite{kea,keb} we could explicitely
show that the currents corresponding to the diagrams in 
Fig. \ref{figcurrent} were conserved {\em separately}. 
This is not the case in inelastic transitions, see Fig. \ref{currcons}. 
The currents alone are far from being conserved separately, 
but the sum of both (solid line) is conserved approximately. The
maximal deviation amounts to less than $30 \, \%$ at  
$-q^2 \approx 0.5 \; \mbox{GeV}^2$. For momentum transfers $-q^2 > 1.5 \;
\mbox{GeV}^2$ the current is found to be nearly conserved. 
While the current conservation of the elastic
currents can be shown analytically in the Mandelstam formalism using 
time and space reversal this is not the case for transition
currents. The violation of the continuity equation indicates that
additional diagrams, as e.g. the coupling
of the photon to the exchanged quark, are needed to fulfill gauge
invariance.   

\section{Summary} \label{secsumm}
We extended our studies of the nucleon in a covariant quark-diquark
model to the 
$\Delta$ resonance. This spin-$\frac{3}{2}$ particle 
is described as a bound state of a quark and a v-diquark with the
Salpeter equation. The interaction kernel is a quark exchange in 
instantaneous approximation. The electromagnetic N-$\Delta$ transitions
are calculated in the Mandelstam formalism from the pseudothreshold
up to $-2 \; \mbox{GeV}^2$. The resulting form factors
are in qualitative agreement with the experimental data, with only 
the dominant magnetic transition $G_M$ coming out too low.
The inclusion of scalar--v-diquark transitions seems to be important
for the nucleon magnetic form factors as well as for the magnetic
N-$\Delta$ transition form factor. We find the correct value for
$E2/M1$ at $q^2=0$. The positive sign of $C2/M1$ contradicts the
experimental findings, its absolute value, however, describes the data
well.  \\
Summarizing, we may state that a pure quark-diquark model in a
covariant approach can account qualitatively for the nucleon form
factors and the N-$\Delta$ transitions up to intermediate momentum
transfers. The
semi-quantitative agreement with experiment is encouraging and may
point to a possible r\^ole of strong quark-quark correlations in
subnuclear physics. \\

{\bf Acknowledgements:} I am grateful to H.R. Petry, B.C. Metsch and 
U.-G. Mei\ss ner for many helpful discussions and W. Pfeil and
R.W. Gothe for very useful comments. This work was supported by the 
Deutsche Forschungsgemeinschaft.

\begin{appendix}
\section{Coupling matrices}
\label{app}
In this section we evaluate the coupling matrices describing 
the coupling of the photon to the v-diquark. We need the 
coupling matrices $\Gamma_{\mu; b a}$ of Sec. III in Ref. \cite{keb}. 
We apply the correct Clebsch-Gordan 
coefficients and take into account the boost prescription of 
Eq. (\ref{rsboost}), thus tearing the boost factors into the 
coupling matrix $\Gamma_{\mu; b a}$:
\bqn \label{boostacoupling}
{\ov \Gamma_{P'}^\Delta}^b(p') \, \Gamma_{\mu; b a} \, \Gamma^N(\vec p\,)^a 
& = & {\ov \Gamma_{(M_\Delta,\vec 0\,)}^\Delta}^{b'}(\vec p_{out}) \, 
S_{\Lambda}^{-1} \, \Lambda^b_{\;\;b'} \, \Gamma_{\mu; b a} \, 
\Gamma^N(\vec p\,)^a \\ 
& =: & {\ov \Gamma_{(M_\Delta,\vec 0\,)}^\Delta}^{b'}(\vec p_{out}) \, 
S_{\Lambda}^{-1} \, \tilde\Gamma_{\mu; b' a} \, \Gamma^N(\vec p\,)^a \; .
\eqn
We then obtain the following one-row coupling matrices in the space
$e_V^{[1]} \otimes (e_0^{[1]},e_V^{[1]})$ (for the notation see
Ref. \cite{keb}):
\bqn
\tilde\Gamma_0 & = & \left( \sqrt{\frac{2}{3}} \, \frac{q}{M_\Delta} \,
  (p_1+p'_1)^0 \;\;,\;\;
\frac{\sqrt{2}}{3} \, \Big( 1-\frac{P'^0}{M_\Delta} \Big) \,
(p_1+p'_1)^0 \right) \\ 
\tilde\Gamma_3 & = & \left( \sqrt{\frac{2}{3}} \, \frac{q}{M_\Delta} \,
  (p_1+p'_1)_3  \;\;,\;\;
\frac{\sqrt{2}}{3} \, \Big( 1-\frac{P'^0}{M_\Delta} \Big) \,
(p_1+p'_1)_3 \right)  \\
\tilde\Gamma_+ & = & \left( 0 \;\;,\;\; -\frac{1}{\sqrt{2}} \, (1+\kappa) \,
  q \, \Big( \frac{1}{3} - \frac{2}{3} \, \frac{P'^0}{M_\Delta}
  \Big) \right) \\
\tilde\Gamma'_+ & = & \left( 0 \;\;,\;\; \frac{1}{\sqrt{6}} \, 
(1+\kappa) \, q \right) \; .
\eqn
We also give the matrices in Lorentz space which appear in the coupling 
to the quark with a v-diquark as spectator:
\bqn
\ov \Gamma_\Delta^{\;b} \, (-g_{b a}) \, \Gamma_N^a & = & 
\tilde{\ov\Gamma}_\Delta^{\;b'} \, S_\Lambda^{-1} \, \Lambda^b_{\;\;b'} \,  
(-g_{b a}) \, \Gamma_N^a \\
& =: & \tilde{\ov\Gamma}_\Delta^{\;b'} \, G_{b' a} \, \Gamma_N^a \; .  
\eqn
With the corresponding Clebsch-Gordan coefficients:
\bqn
G^{\mu=0,3} & = & \left( \sqrt{\frac{2}{3}} \, \frac{q}{M_\Delta} \;\;,\;\;
\frac{\sqrt{2}}{3} \, \Big( 1-\frac{P'^0}{M_\Delta} \Big) \right) \\
G^{\mu=+} & = & \left( \sqrt{\frac{2}{3}} \, \frac{q}{M_\Delta} \;\;,\;\;
\frac{\sqrt{2}}{3} \, \frac{P'^0}{M_\Delta} \right) \\
G'^{\mu=+} & = & \left( 0 \;\;,\;\; \sqrt{\frac{2}{3}} \right) \; .
\eqn
\end{appendix}

\begin{table}[t]
\begin{center}
\begin{tabular}{c|ccccccc}
& $m_q$ & $m_s = m_v$ & $g^N$ & $g_v^\Delta$ & $\lambda$ & $\kappa$ & 
$\kappa_{sv}$ \\  
\hline
Set A &
440 MeV/c$^2$ & 
800 MeV/c$^2$ & 
17.76 &
8.50 &
0.30 fm &
1.1 & - \\
Set B &
440 MeV/c$^2$ & 
800 MeV/c$^2$ & 
17.76 &
8.50 &
0.30 fm &
-0.07 & 2.4  
\end{tabular}
\caption{The parameters of the model: Set A and Set B.}
\label{deltaparam}
\end{center}
\end{table}
\begin{table}
\begin{center}
\begin{tabular}{c|cccccc}
 & $\sqrt{\langle r^2 \rangle_E^p}$ & $\langle r^2 \rangle_E^n$ & 
$\sqrt{\langle r^2 \rangle_M^p}$ &
$\sqrt{\langle r^2 \rangle_M^n}$ &
$\mu_p$ &
$\mu_n$ \\ 
\hline
Set A & 0.79 fm  & -0.110 fm$^2$ & 
0.72 fm  & 0.86 fm  & 
2.45 $\mu_N$ & -1.42 $\mu_N$ \\ 
Set B & 0.79 fm  & -0.110 fm$^2$ & 
0.74 fm  & 0.75 fm &
2.44 $\mu_N$ & -1.91 $\mu_N$ \\ 
exp. & 0.847 fm & -0.119 fm$^2$ &
0.836 fm & 0.889 fm & 
2.793 $\mu_N$ & -1.913 $\mu_N$   
\end{tabular}
\caption{Static nucleon properties as they
  result from the threshold behaviour of the electromagnetic 
  nucleon form factors. For the experimental data see the analysis of 
  Ref. \protect\cite{ulf}.}  
\label{tabresults}
\end{center}
\end{table}


\begin{thebibliography}{99}
\bibitem{baryon} Proc. Int. Conf. 'Baryons '95', Santa F\'e (1995)
\bibitem{drechsel} O. Hanstein, D. Drechsel and L. Tiator,
  nucl-th/9605008, submitted to Phys. Lett. B
\bibitem{bernstein} A.M. Bernstein, S. Nozawa and M.A. Moinester, 
Phy. Rev. C {\bf 47}, 1247 (1993)
\bibitem{kea} V. Keiner, Z. Phys. A {\bf 354}, 87 (1996) 
\bibitem{keb} V. Keiner, hep-ph/9603226, to be published in
  Phys. Rev. C {\bf 54} (1996)
\bibitem{anselmino} M. Anselmino, E. Predazzi, S. Ekelin,
  S. Fredriksson and D.B. Lichtenberg, Rev. Mod. Phys., Vol. {\bf 65},
  No.4, 1199 (1993) 
\bibitem{leinweber} D.B. Leinweber, Phys. Rev. D {\bf 47}, 5096 (1993) 
\bibitem{kroll} P. Kroll, M. Sch\"urmann and W. Schweiger, Z. Phys. A
  {\bf 342}, 429 (1992)
\bibitem{reinhardt} H. Reinhardt, Phys. Lett. B {\bf 244}, 316 (1990)
\bibitem{buck} A. Buck, R. Alkofer and H. Reinhardt, Phys. Lett. B
  {\bf 286}, 29 (1992) 
\bibitem{huang} S. Huang and J. Tjon, Phys. Rev. C {\bf 49}, 1702 (1994)
\bibitem{meyer} H. Meyer, Phys. Lett. B {\bf 337}, 37 (1994)
\bibitem{anselkroll} M. Anselmino, P. Kroll and B. Pire, Z. Phys. C {\bf
  36}, 89 (1987) 
\bibitem{bethe} E.E. Salpeter and H.A. Bethe, Phys. Rev. {\bf 84},
  132 (1951) 
\bibitem{ishii} N. Ishii, W. Bentz and K. Yazaki, Phys. Lett. B {\bf 318}, 26
  (1993)
\bibitem{mandelstam} S. Mandelstam, Proc. Roy. Soc. {\bf 233}, 248 (1955)
\bibitem{jones} H.F. Jones and M.D. Scadron, Ann. Phys. {\bf 81}, 1
  (1973)   
\bibitem{devenish} R.C.E. Devenish, T.S. Eisenschitz and
  J.G. K\"orner, Phys. Rev. D {\bf 14}, 3063 (1976)
\bibitem{weber} H.J. Weber, Ann. Phys. {\bf 207}, 417 (1991)
\bibitem{hanhart} C. Hanhart and S. Krewald, Phys. Lett. B {\bf 344}, 55 (1995)
\bibitem{bartel} W. Bartel, B. Dudelzak, H. Krehbiel, J. McElroy,
  U. Meyer-Berkhout, W. Schmidt, V. Walther and G. Weber,
  Phys. Lett. B {\bf 28}, 148 (1968)  
\bibitem{stein} S. Stein, W.B. Atwood, E.D. Bloom, R.L.A. Cottrell, H.
  DeStaebler, C.L. Jordan, H.G. Piel, C.Y. Prescott, R. Siemann and
  R.E. Taylor, Phys. Rev. D {\bf 12}, 1884 (1975)
\bibitem{ash} W.W. Ash, K. Berkelman, C.A. Lichtenstein,
  A. Ramanauskas and R.H. Siemann, Phys. Lett B {\bf 24}, 165 (1967) 
\bibitem{alder} J.C. Alder {\em et al.}, Nucl. Phys. {\bf B46},
  573 (1972) 
\bibitem{siddle} R. Siddle {\em et al.}, Nucl. Phys. {\bf B35},
  93 (1971)
\bibitem{beck} R. Beck {\em et al.} (MAMI), submitted to Phys. Rev. Lett.,
  Sept. 1996
\bibitem{crawford} R.L. Crawford, Nucl. Phys. {\bf B28}, 573
  (1971); exp. data from C. Mistretta {\em et al.}, Phys. Rev. {\bf
  184}, 1487 (1969) 
\bibitem{gothe} F. Kalleicher, Ph.D. thesis, Mainz University (1993); 
D. Jakob, Ph.D. thesis, Bonn University (1996); 
R.W. Gothe {\em et al.}, ELSA proposal (1995)
\bibitem{cardarelli} F. Cardarelli, E. Pace, G. Salm\`e and S. Simula,
  Phys. Lett. B {\bf 371}, 7 (1996)
\bibitem{ulf} P. Mergell, U.-G. Mei\ss ner and D. Drechsel,
  Nucl. Phys. {\bf A596}, 367 (1996)
\end{thebibliography}
\end{document}